\documentclass[aps,reprint,prl,showpacs,amsmath,amssymb, nobibnotes]{revtex4-1}
\usepackage{hyperref}
\usepackage{epsfig}
\usepackage{graphicx}
\usepackage{amssymb,amsmath}
\usepackage{fancybox,framed}
\usepackage{dsfont}
\usepackage{mathtools}
\usepackage{braket}
\usepackage{slashed}
\usepackage{rotating}
\usepackage{bbold,amsfonts}
\newcommand{\dota}{\dot{\alpha}}
\newcommand{\dotb}{\dot{\beta}}
\renewcommand{\a}{\alpha}
\renewcommand{\b}{\beta}

\renewcommand{\d}{\delta}
\newcommand{\pa}{\partial}
\newcommand{\g}{\gamma}
\newcommand{\G}{\Gamma}

\newcommand{\D}{\Delta}
\newcommand{\e}{\epsilon}

\renewcommand{\l}{\lambda}
\renewcommand{\L}{\Lambda}

\newcommand{\s}{\sigma}

\renewcommand{\t}{\tau}

\newcommand{\Tr}{\textup{Tr}}

\newcommand{\beq}{\begin{equation}}
\newcommand{\eeq}{\end{equation}}

\newcommand{\ie}{\emph{i.e.}}
\newcommand{\ii}{\mathrm{i}}

\begin{document}

\title{Line defects and radiation in $\mathcal{N}=2$ theories}%

\author{Lorenzo Bianchi,$^{1}$ Madalena Lemos,$^{2}$ and Marco Meineri$^{3}$}%
\affiliation{$^{1}$ Center for Research in String Theory - School of Physics and Astronomy Queen Mary University of London, Mile End Road, London E1 4NS, UK\\
$^{2}$ DESY Hamburg, Theory Group, Notkestrasse 85, D-22607 Hamburg, Germany\\
$^{3}$ Institute of Physics, \'{E}cole Polytechnique F\'{e}d\'{e}rale de Lausanne (EPFL),
Rte de la Sorge, BSP 728, CH-1015 Lausanne, Switzerland}

\begin{abstract}
We study the conformal data of a generic superconformal half-BPS line defect in a four-dimensional $\mathcal{N} = 2$ theory. We prove a theory independent relation between the one-point function of the stress tensor in the presence of the line defect and the two-point function of the displacement operator.
When the defect is interpreted as a heavy charged particle in a gauge theory, the result relates the energy emitted through Bremsstrahlung with the coupling of the stress tensor to the particle at rest.
\end{abstract}

\pacs{;\qquad Preprint numbers: DESY 18-071 }

\maketitle

\section{Introduction}

When studying a Quantum Field Theory, the typical experiment consists in probing the vacuum and measuring the response of the system far from the probed region. A prototypical observable is the energy emitted by a heavy particle which accelerates in a medium that contains massless degrees of freedom. In a gauge theory, this process is called Bremsstrahlung, and the natural question to be asked is the relation between the energy emitted and the trajectory of the particle. The answer, in any conformal field theory (CFT), and for any one-dimensional classical probe -- or conformal line defect for short -- is \cite{Correa:2012at}
\beq
 \D E=2\pi B \int\! dt\, \dot v^2\,,
 \label{brem}
\eeq
where $\dot v$ is the proper acceleration and $B$ is the Bremsstrahlung function. For instance, in Maxwell's theory, $B=e^2/12\pi^2$, where $e$ is the charge of the accelerated particle. In \cite{Correa:2012at}, an exact formula was given for the case of a 1/2-BPS Wilson line in $\mathcal{N}=4$ SYM: $B= \l \partial_\l \braket{W}/2\pi^2$, where $\l$ is the 't Hooft coupling and $\braket{W}$ is the expectation value of the circular Wilson loop, which can be computed exactly \cite{Erickson:2000af,Drukker:2000rr,Pestun:2007rz}. In fact, $B$ is accessible through a variety of physical observables. Especially relevant is its relation to the cusp anomalous dimension $\G_{\text{cusp}}$, which controls the logarithmic divergence in the expectation value of a Wilson loop with a cusp, as well as the leading IR divergence in the scattering of massive particles \cite{Polyakov:1980ca,Korchemsky:1987wg}. If $\phi$ is the angle of the cusp, $B$ determines the smooth limit $\phi\to 0$,  $\G_{\text{cusp}}(\phi)\sim-B \phi^2$.

On the other hand, the energy deposited at infinity is computed by integrating the appropriate component of the stress tensor. In particular, in the presence of a constantly accelerated probe, the expectation value of the stress tensor is determined by a single constant, usually dubbed $h$ in this context. One might expect that a theory independent relation should exist between $B$ -- the emitted energy -- and $h$ -- the energy measured at infinity. Unfortunately, this is not the case, probably due to non radiative contributions to $h$. In \cite{Lewkowycz:2013laa}, a way of subtracting these spurious terms was devised for BPS line defects in theories with enough supersymmetries, leading to the conjecture
\beq
B= 3\,h\,,
\label{conj}
\eeq
in four dimensions. Direct computation of the two sides confirmed eq. \eqref{conj} for 1/2-BPS Wilson lines in $\mathcal{N}=4$ SYM \cite{Okuyama:2006jc,Gomis:2008qa,Correa:2012at}, while evidence has been provided for 1/2-BPS Wilson loops in $\mathcal{N}=2$ theories \cite{Fiol:2015spa}, as well as for the three-dimensional version of the conjecture for 1/2 and 1/6-BPS Wilson lines in ABJM \cite{Lewkowycz:2013laa,Bianchi:2017svd,Bianchi:2017ozk,Bianchi:2018scb,Bianchi:2018bke}. The relation of the stress tensor with a change in the geometry motivated the authors of \cite{Fiol:2015spa} to also conjecture an exact expression for the Bremsstrahlung function $B=\partial_b \log \braket{W_b}|_{b=1}/4\pi^2$, where the Wilson loop is placed on a squashed sphere with squashing parameter $b$. 

We see that the relation \eqref{conj} not only has a clear and interesting physical content, it also guarantees the computational advantage of accessing the same quantity through different observables. The goal of this paper is to prove that eq. \eqref{conj} holds for any 1/2-BPS line defect in four-dimensional $\mathcal{N}=2$ theories. 

The main ingredient is the fact that every conformal defect supports a displacement operator $\mathbb{D}$, which expresses the response of the theory to a deformation of the defect. The coefficient of its two-point function is proportional to $B$ \cite{Correa:2012at}. Crucially, the two-point function of $\mathbb{D}$ with the stress tensor contains both $B$ and $h$ \cite{Billo:2016cpy}. In the rest of the paper, we derive the constraints imposed by supersymmetry on the coupling of the stress tensor multiplet to the displacement multiplet, and we show that they are sufficient to prove the conjecture \eqref{conj} in the most general formulation.

\section{1/2-BPS Wilson lines}
Although our result is general, let us describe the most important example. The amount of supersymmetry preserved by a Wilson line is determined by the contour along which the line is stretched and by the explicit form of the connection. For $\mathcal{N}=2$ theories in four dimensions it is possible to define a Wilson line which preserves half of the original supersymmetries. In particular, for a straight line contour parametrized as $x^{\mu}(\tau)=(0,0,0,\tau)$ in Euclidean signature, the supersymmetry variation of the gauge field can be compensated by the transformation of the complex scalar in the vector multiplet such that the operator
\begin{align}
 W&=\Tr\mathcal{P} \exp\left[ \ii \int d\tau \mathcal{A}\right]  & \mathcal{A}&=A_{4} + \phi + \bar \phi\,,
\end{align}
is annihilated by the combination of supercharges
\begin{equation}
 \mathcal{Q}^a_{\a}=Q^a_{\a}-\ii \s^4_{\a \dot\a } \bar Q^{a\dot \a}\,.
\end{equation}
Here and in the following, $\alpha=1,2$ is a Lorentz index, $a=1,2$ is a R-symmetry index, and $\s^\mu = (\tau^i,\ii \mathbb{1})$, $\tau^i$ being the Pauli matrices. Raising and lowering conventions are as in Wess and Bagger, but $\e_{12}=-\e^{12}=1$.
The set of preserved generators also includes the one-dimensional conformal algebra $\{P_4,D,K_4\}$, the rotations in the orthogonal directions $M^{\a}{}_\b$, the superconformal charges $\mathcal{S}^a_{\a}=S^a_{\a}+\ii\s^4_{\a\dot\a} \bar S^{a \dot \a}$ and the preserved $SU(2)_R$ R-symmetry $J^a{}_b$. These generators span a $osp(4^*|2)$ subalgebra of the full $su(2,2|2)$ superalgebra. Defect operators are organized in irreducible representations of the preserved superalgebra. In particular highest weight operators are characterized by three charges $[\D,j,R]$, associated to the maximal bosonic subalgebra $so(1,2)\oplus su(2) \oplus su(2)_R$. 

In the following we will not rely on a Lagrangian description of the line defect, rather we only use the preserved superalgebra. The results are thus valid for any $4d$ $\mathcal{N}\geq 2$ line defect that preserves $osp(4^*|2)$.

\section{Stress tensor and displacement supermultiplets}
It is a well known fact of supersymmetric theories that a single multiplet accommodates the stress tensor operator as well as the Noether currents associated to supersymmetry and R-symmetry. In $\mathcal{N}=2$ theories in four dimensions, the superprimary of the current multiplet is a scalar $O_2$. The multiplet includes two fermions $\chi_\a^a,\, \bar{\chi}^{\dota}_a$, the R-symmetry currents $j^{\mu}$ and $j^{\mu}{}^a{}_b$ associated to $U(1)_R$ and $SU(2)_R$ respectively, the supercurrents $J^{\mu a}_{\a}$ and $\bar J^{\mu\dot \a}_a$ associated to the supercharges $Q^a_\a$ and $\bar Q^{\dot \a}_a$, and the $(1,0)$ and $(0,1)$ fields $H_{\a\b},\, \bar{H}_{\dota\dotb}$. When the theory is perturbed by an extended probe, some of the currents are broken by contact terms localized on the defect. For instance, the stress tensor conservation law in the presence of a straight line defect is modified as follows:
\begin{equation}\label{eq:disp}
 \pa_\mu T^{\mu m}(x,\tau)=\d^3(x) \mathbb{D}^m(\tau)\,,
\end{equation}
where the index $\mu=1,...,4$ spans the whole four dimensional space, while the index $m=1,2,3$ labels the directions orthogonal to the line, which is stretched along $x^4$. The defect operator on the r.h.s. of \eqref{eq:disp} is called displacement operator and has quantum numbers $[2,1,0]$. Similarly, the conservation law for the broken $U(1)$ R-symmetry current reads
\begin{equation}\label{eq:Rop}
 \pa_\mu j^{\mu}(x,\tau)=\d^3(x) \mathbb{O}(\tau)\,,
\end{equation}
where $\mathbb{O}(\tau)$ is a defect scalar operator of charges $[1,0,0]$. Also the supersymmetry currents $J^{\mu a}_{\a}$ and $\bar J^{\mu\dot \a}_a$ split into preserved ($\mathcal{J}^{\mu a}_{\a}$) and broken ($\mathfrak{J}^{\mu a}_{\a}$) supercurrents. For the latter the conservation law is broken to
\begin{equation}\label{eq:lambda}
 \partial_{\mu}\mathfrak{J}^{\mu a}_{\a}=\d^3(x) \mathbb{\L}^a_{\a} (\tau)\,.
\end{equation}
In this case the fermionic operator on the r.h.s. has charges $[3/2,1/2,1/2]$. The algebraic structure of $osp(4^*|2)$ forces the three defect operators $\mathbb{O}$, $\mathbb{\L}^a_{\a}$ and $\mathbb{D}^m$ to sit in the same supermultiplet. This can be seen in the following way. Consider the Hilbert space on a cylinder surrounding the Wilson line. States are defined by Euclidean path-integration, in particular $\ket{W}$ is obtained by path-integrating with no insertions other than the straight line defect. Eqs. \eqref{eq:disp} to \eqref{eq:lambda} imply that integrating the broken currents on the cylinder produces the same state as the integration of the corresponding defect operators:
\begin{align}
 \mathfrak{R}\ket{W}&=-\ii \int d\tau\, \mathbb{O}(\tau) \ket{W}\equiv-\ii \int d\tau \ket{\mathbb{O}}\,,\\
 \mathfrak{Q}^a_{\a}\ket{W}&=-2\ii\int d\tau\, \mathbb{\L}^a_{\a}(\tau) \ket{W}\equiv -2\ii\int d\tau \ket{\mathbb{\L}^a_{\a}}\,,\\
 \mathfrak{P}_{\a\b}\ket{W}&=-\ii\int d\tau\, \mathbb{D}_{\a\b}(\tau) \ket{W}\equiv-\ii \int d\tau \ket{\mathbb{D}_{\a\b}}\,.
\end{align}
Here $\mathfrak{R},\ \mathfrak{Q}^a_{\a},\ \mathfrak{P}_{\a\b}$ are obtained by integrating $ j^{\mu},\ \mathfrak{J}^{\mu a}_{\a},\ T^{\mu m}$ respectively. 
By repeatedly applying the commutation relations for the $osp(4^*|2)$ algebra one finds
\begin{align}
 \mathcal{Q}^a_{\a}\mathfrak{R}\ket{W}=  \tfrac12 \mathfrak{Q}^a_{\a} \ket{W}&= -\ii\int d\tau \ket{\mathbb{\L}^a_{\a}}\,,\nonumber\\
 \mathcal{Q}^a_{\a}\mathfrak{Q}^b_{\b}\ket{W}=2 \e^{ab}\mathfrak{P}_{\a\b} \ket{W}&= -2 \ii \e^{ab} \int d\tau \ket{\mathbb{D}_{\a\b}}\,, \nonumber \\
 \mathcal{Q}^a_\a\mathfrak{P}_{\b\g} \ket{W}&=0 \,. \label{disptop}
\end{align}
Notice that this procedure is blind to conformal descendants which would appear as total derivatives in the integrand. Nevertheless, the coefficient of the total derivative can always be fixed by implementing the action of the anti-commutator $\{\mathcal{Q}^a_{\a},\mathcal{Q}^b_{\b}\}$. After doing that we find
\begin{align}
 &\mathcal{Q}^a_{\a}\mathbb{O}=\mathbb{\L}^a_{\a}\,, \qquad \mathcal{Q}^a_{\a} \mathbb{\L}^b_{\b}=2\e^{ab}\mathbb{D}_{\a\b}-2\e^{ab} \e_{\a\b} \pa_{\tau} \mathbb{O}\,, \nonumber \\
   &\mathcal{Q}^a_{\d} \mathbb{D}_{\a\b} = -  \pa_{\tau} \mathbb{\L}^a_{\a} \epsilon_{\b \d}-  \pa_{\tau} \mathbb{\L}^a_{\b} \epsilon_{\a \d}\,. 
   \label{dispsusy}
\end{align}
Equation \eqref{disptop} is a manifestation of the general fact that the displacement operator is always the top component of its protected supermultiplet. This is often a sufficient condition to identify the displacement supermultiplet only based on representation theory. Also in this case a careful analysis of the $osp(4^*|2)$ representation theory shows that the multiplet we described is the only multiplet which can accommodate the displacement operator, \ie, a top component of dimension one, singlet under $SU(2)_R$ and a vector of $SO(3)$. In particular, the highest weight must obey the semi-shortening $ \mathcal{Q}^1_{\a} \mathcal{Q}^1_{\b}\epsilon^{\a \b} \left|\psi\right.\rangle = 0$, thus fixing its scaling dimension as $\Delta=2R+j+1$. Further shortening of the multiplet comes from the fact that the superconformal primary has $R=j=0$, which then matches the multiplet displayed in \eqref{dispsusy}.

For the particular case of the Wilson line, the previous arguments can be made very explicit by using the variation of the Wilson line under an infinitesimal symmetry transformation
\begin{align}
 \d \braket{W}=\ii\int d\tau \braket{\d\mathcal{A}}_W\,,
\end{align}
which immediately allows to compute the operators in the displacement supermultiplet
\begin{align}
\mathbb{O}&=
%\ii \d_{\mathfrak{R}} \mathcal{A}=
\bar \phi-\phi\,,\\
\mathbb{\L}^a_{\a}&=
%\ii \d_{\mathfrak{Q}^a_\a} \mathcal{A}=
\l^a_\a-\ii \s_{\a\dot\a}^4 \bar \l^{a\dot\a}\,,\\
\mathbb{D}^m&=
%\ii \d_{\mathfrak{P}_m} \mathcal{A}=
-\ii(F^{m4}+D^m\phi+D^m\bar \phi)\,.
\end{align}
Using the supersymmetry transformations given in \cite{Fiol:2015spa}, one can recover equation \eqref{dispsusy}, with the important observation that the role of the total derivative for the Wilson line defect is played by the covariant derivative $\mathcal{D}_4=\pa_4- \ii \mathcal{A}$ (see \cite{Cooke:2017qgm,Bianchi:2017ozk,Bianchi:2018scb}).

\section{Displacement two-point functions}
The normalization of the operators in the displacement supermultiplet is fixed by the Ward identities \eqref{eq:disp}, \eqref{eq:Rop} and \eqref{eq:lambda}. Therefore the Zamolodchikov norms defined by the defect two-point functions are defect conformal data. In particular, implementing superconformal Ward identities for the preserved supercharges, one can fix all the defect two-point functions of the operators in the displacement supermultiplet in terms of a single constant. Such constant can be defined by the displacement two-point function
\begin{equation}
 \braket{\mathbb{D}^m(0) \mathbb{D}^n(\tau)}_W=\frac{12 B\, \d^{mn}}{\t^4}\,,
\end{equation}
where we used the result of \cite{Correa:2012at} to relate the displacement two-point function to the Bremsstrahlung function.
Then, using superconformal Ward identities, one can fix
\begin{align}
 \braket{\mathbb{O}(0)\mathbb{O}(\tau)}_W&=\frac{2B}{ \tau^2}\,,\\
  \braket{\mathbb{\L}^a_\a(0)\mathbb{\L}^b_{\b}(\tau)}_W&=-\e_{\a\b}\e^{ab} \frac{8 B}{ \tau^3}\,.
\end{align}

\section{Stress tensor one-point functions}
In this section we briefly review the results of \cite{Fiol:2015spa}, where the authors showed that non-vanishing one-point functions of the operators in the stress tensor multiplet can be fixed up to a single constant. Once more we define such constant using the top component of the multiplet, \ie, the stress tensor
\begin{align}
 \braket{T_{mn}(x,\tau)}&=-\frac{h}{x^4} (\d_{mn}-2n_m n_m)\,, \\
 \braket{T_{m4}(x,\tau)}&=0\,, \qquad \braket{T_{44}(x,\tau)}_W=\frac{h}{x^4}\,,
\end{align}
where $n_m = \tfrac{x_m}{|x|}$.
The other operators which acquire a non-vanishing one-point function in presence of a 1/2-BPS line defect are the scalar superprimary $O_2$ and the two-form $H_{\a}{}^{\b}$
\begin{align}
 \braket{O_2(x,\tau)}_W&=\frac{3\, h}{8 x^2}\,,\\
  \braket{H_{\a}{}^\b(x,\tau)}_W&=\frac{3 \ii \,  h\, x_m (\sigma^m)_{\a}{}^{\b}}{4 x^4}\,,
\end{align}
where the $3d$ sigma matrices $(\sigma^m)_{\a}{}^{\b}$ are  taken to be the Pauli matrices.

\section{Displacement stress tensor two-point function}

The authors of \cite{Billo:2016cpy} showed that the bulk to defect two-point function coupling the stress tensor to the displacement operator can be completely fixed in terms of $B$ and $h$. In particular, the residual conformal symmetry allows for three different structures with the correct transformation properties. These structures are associated to three independent constants, which are then related to $B$ by the Ward identity
\begin{equation}
 \pa_{\mu} \braket{T^{\mu m}(x,\tau)\mathbb{D}^n(\tau')}_W=\d^3(x) \braket{\mathbb{D}^m(\tau) \mathbb{D}^n(\tau')}_W\,,
\end{equation}
and to $h$ by the integrated relation
\begin{equation}
 \braket{\mathfrak{P}^m T^{\mu\nu}(x,\tau)}_W=-\ii\int d\tau' \braket{T^{\mu \nu}(x,\tau)\mathbb{D}^m(\tau')}_W\,.
\end{equation}
In the following we will apply the same procedure to the other components of the stress tensor and displacement supermultiplets. For the defect superprimary $\mathbb{O}$ and a generic bulk operator $\mathcal{O}$ we can use
\begin{equation}\label{U1ward}
  \braket{ \mathfrak{R} \mathcal{O}(x,0)}_W=-\ii\int d\tau \braket{\mathcal{O}(x,0)\mathbb{O}(\tau)}_W\,,
\end{equation}
which is a rather powerful constraint since it sets to zero the coupling of $\mathbb{O}$ with $U(1)$ neutral operators (an exception to this rule is the two-point function $\braket{j^\mu \mathbb{O}}_W$ which we will consider below). Since the stress tensor supermultiplet contains a single $U(1)$ charged bosonic operator, equation  \eqref{U1ward} allows to fix
\begin{equation}\label{HO}
 \braket{H_{\a}{}^{\b}(x,0) \mathbb{O}(\tau)}_W=-\frac{3\, h\,  n_m (\sigma^m)_{\a}{}^{\b}}{4\pi x^2(x^2+\tau^2)}\,.
\end{equation}
As we mentioned, the only other non-vanishing two-point function involving $\mathbb{O}$ is the one with the $U(1)$ current $j^\mu$, for which the r.h.s. of \eqref{U1ward} is identically zero (one can easily see that the kinematical structure derived in \cite{Billo:2016cpy} integrates to zero). Nevertheless, we can use the additional Ward identity
\begin{equation}
 \pa_\mu \braket{j^{\mu}(x, \tau) \mathbb{O}(\tau')}_W=\d^3(x) \braket{\mathbb{O}(\tau) \mathbb{O}(\tau')}_W\,,
\end{equation}
to determine the two-point function in term of $B$
\begin{align}
  \braket{j^4(x,0)\mathbb{O}(\tau)}_W&=\frac{B}{\pi} \frac{\tau}{|x|(x^2+\tau^2)^2}\,,  \label{jO1}\\
 \braket{j^m(x,0) \mathbb{O}(\tau)}_W&=\frac{B}{2\pi}\frac{ n^m (\tau^2-x^2)}{x^2(x^2+\tau^2)^2}\,. \label{jO2}
\end{align}

Similar Ward identities can be derived using broken supercharges
\begin{align}\label{superWardbroken}
  \braket{\mathfrak{Q}_{\a}^a \mathcal{X}(x,0)}_W=-2\ii \int d\tau \braket{\mathcal{X}(x,0)\mathbb{\L}_{\a}^a(\tau)}_W\,,
\end{align}
where $\mathcal{X}$ is a bulk fermionic operator. The latter constraint is particularly powerful since it relates fermionic bulk to defect correlators to bosonic bulk one-point functions. Since for the two-point function $\braket{\chi_{\a}^a(x,0) \mathbb{\L}^{\b}_b(\tau)}_W$ conformal invariance together with parity invariance allow for a single kinematical structure, the relation \eqref{superWardbroken} implies
\begin{align}\label{chiL}
 \braket{\chi_{\a}^a(x,0) \mathbb{\L}^{\b}_b(\tau)}_W &= \frac{3h}{2\pi} \frac{\ii\,\d_{\a}{}^{\b} \tau-(\s_m)_{\a}{}^{\b} x^m}{|x|(x^2+\tau^2)^2}\,.
\end{align}

\section{Supersymmetric Ward identities}

We can obtain further constraints on the correlator between the displacement and stress tensor supermultiplets by applying supersymmetric Ward identities which come from preserved supercharges. For instance, given the two-point function $\braket{ \chi^a_\a(x,0)\mathbb{O}(\tau)}_W$, the constraint $\d_{\text{susy}} W=0$  translates into $\braket{\d_{\text{susy}}[\chi^a_\a(x,0)\mathbb{O}(\tau)]}_W=0$ and consequently
\begin{align}
 & 0=\d^a_b\braket{{H_{\a}}^\b(x,0)\mathbb{O}(\tau)}_W+\d^a_b\tfrac{1}{2}\braket{j_\a{}^{\b}(x,0),\mathbb{O}(\tau)}_W \nonumber\\
 & +\tfrac{\ii}{2}\d^a_b \d_\a^\b \braket{j_4(x,0),\mathbb{O}(\tau)}_W -\braket{\chi^a_\a(x,0)\mathbb{\L}_b^\b(\tau)}_W \,,\label{Wardid}
\end{align}
where $j_\a{}^{\b}=j_m (\s^m)_\a{}^{\b}$. Inserting equations \eqref{HO}, \eqref{jO1}, \eqref{jO2} and \eqref{chiL} into equation \eqref{Wardid} one can easily see that the only solution is
\begin{align}
 B=3\, h\,,
\end{align}
thus proving the relation conjectured in \cite{Lewkowycz:2013laa,Fiol:2015spa}.

\section{Outlook}

In this paper, we proved a theory independent relation between energy emitted by a supersymmetric line defect and the value of the stress tensor in the background of the line. From a defect CFT point of view, eq. \eqref{conj} may be a precious input in the study of this class of line defects through the conformal bootstrap \cite{Liendo:2012hy,Gliozzi:2015qsa}. Notice that relations between $h$ and the two-point function of the displacement are not uncommon in the realm of defect CFTs \cite{Bianchi:2015liz,Bianchi:2016xvf,Balakrishnan:2017bjg,Dong:2016wcf,Balakrishnan:2016ttg,Herzog:2017xha}. Eq. \eqref{conj} essentially follows from the representation theory of $osp(4^*|2)$ and from the transformation rules of conformal defects under the symmetries of the theory they belong to, summarized in \eqref{eq:disp}-\eqref{eq:lambda}. At a technical level, it would be nice to set up a more concise derivation in superspace. In this respect, let us remark that in the special case of 1/2-BPS defects in four-dimensional $\mathcal{N}=4$ theories, eq. \eqref{conj} can be quickly derived from the formulation developed in \cite{Liendo:2016ymz}\footnote{It is worth mentioning that the interpretation of the Wilson line as a superconformal defect in $\mathcal{N}=4$ SYM has received a renewed interest recently \cite{Giombi:2017cqn,Cooke:2017qgm,Giombi:2018qox}}. It would also be interesting to study the detailed contribution of the self-energy to $h$ in the picture of \cite{Lewkowycz:2013laa}: this may also boost our understanding of non supersymmetric Wilson lines (see also \cite{Beccaria:2018ocq,Beccaria:2017rbe} for related discussions). Of course, the actual value of $B$ and $h$ is interesting per se. The prescription given in \cite{Fiol:2015spa} for computing $h$ using the matrix model for a deformed background, though motivated by a clear geometrical picture, surely deserves further analysis. In particular, it would be interesting to understand whether a general prescription exists for computing correlation functions of current multiplet operators by taking derivatives with respect to the squashing parameter of the matrix model \cite{Billo:2017glv,Billo:2018oog}. Finally, an immediate direction of future research concerns other supersymmetric defects: the three-dimensional counterpart of the conjecture \cite{Lewkowycz:2013laa} may be tackled with the same techniques, and it is interesting to ask more generally if relations of this kind are obeyed by defects of different dimension and codimension.

\begin{acknowledgments}
It is a pleasure to thank Marco Bill\`o, Francesco Galvagno, Alberto Lerda and Jo\~{a}o Penedones for very useful discussions. The work of LB is supported by the European Union’s Horizon 2020 research and innovation programme under the Marie Sklodowska-Curie grant agreement No 749909. MM is supported by the Simons Foundation grant 488649 (Simons collaboration on the non-perturbative bootstrap) and by the National Centre of Competence in Research SwissMAP funded by the Swiss National Science Foundation. ML was partially supported  by the German Research Foundation (DFG) via the Emmy Noether program ``Exact results in Gauge theories''.
\end{acknowledgments}

\bibliography{biblio.bib}
\end{document}